\documentclass[aps,prl,twocolumn,floats,epsfig,showpacs]{revtex4-1}

\usepackage{bbm}
\usepackage{amssymb}
\usepackage{ifpdf}
\usepackage{color}
\usepackage{graphicx,epsfig}

\begin{document}

\title{Nematic Confined Phases in the $U(1)$ Quantum Link Model on a Triangular
Lattice: \\ 
An Opportunity for Near-Term Quantum Computations of String Dynamics on a Chip}

\author{D.\ Banerjee$^1$, S.\ Caspar$^2$, F.-J.\ Jiang$^3$, J.-H.\ Peng$^3$, 
and U.-J.\ Wiese$^4$}

\affiliation{$^1$Saha Institute of Nuclear Physics, HBNI, 1/AF Bidhannagar, 
Kolkata 700064, India \\
$^2$ InQubator for Quantum Simulation, Department of Physics, 
University of Washington, Seattle, WA 98195 \\
$^3$ Department of Physics, National Taiwan Normal University
88, Sec.\ 4, Ting-Chou Rd., Taipei 116, Taiwan \\
$^4$Albert Einstein Center, Institute for Theoretical Physics, 
University of Bern, 3012 Bern, Switzerland}
\begin{abstract}

The $U(1)$ quantum link model on the triangular lattice has two 
rotation-symmetry-breaking nematic confined phases. Static external charges are 
connected by confining strings consisting of individual strands with 
fractionalized electric flux. The two phases are separated by a weak first 
order phase transition with an emergent almost exact $SO(2)$ symmetry. We 
construct a quantum circuit on a chip to facilitate near-term quantum 
computations of the non-trivial string dynamics.

\end{abstract}

\maketitle

The confinement of quarks and gluons inside hadrons is a central dynamical
mechanism in Quantum Chromodynamics (QCD). In a pure Yang-Mills theory, in which
quarks appear only as static external color charges, quarks and anti-quarks are
connected by unbreakable confining strings. At large distances $r$ the static
quark-anti-quark potential $V(r) \sim \sigma r$ is dominated by the string
tension $\sigma$. The strings themselves are interesting dynamical objects that 
support massless excitations, which are described by a systematic low-energy 
effective theory of Goldstone bosons \cite{Lue80,Lue81,Lue81a}. This low-energy 
effective string theory predicts the universal sub-leading L\"uscher term 
correction to $V(r)$. In the presence of a static quark-anti-quark pair some 
spatial symmetries as well as charge conjugation are explicitly broken. The 
string excitations can be classified by the irreducible representations of the 
remaining unbroken subgroup, and are again predicted by the effective string
theory. The dynamics of the string have been studied in great detail by Monte 
Carlo simulations in the framework of Wilson's lattice gauge theory 
\cite{Lue01,Lue02,Jug03,Maj03,Lue04,Bra09,Gli10,Bra16} and quantitative 
agreement with the low-energy effective theory has been established. 

Quantum link models \cite{Hor81,Orl90,Cha97,Bro99,Bro04} provide a 
generalization of Wilson's framework of lattice gauge theory \cite{Wil74}. In 
contrast to the Wilson theory, quantum link models have a finite-dimensional 
link Hilbert space, while still maintaining exact gauge symmetry. Quantum link 
models capture a wider range of physical phenomena than those that are 
accessible in the Wilson framework. This includes ``crystalline confined 
phases'' \cite{Ban14} which are characterized by the spontaneous breakdown of 
lattice translation symmetry as well as the splitting of confining strings 
into individual strands that carry fractionalized electric flux, both in 
Abelian \cite{Ban13} and in non-Abelian \cite{Ban18} quantum link models. The
$(2+1)$-d $U(1)$ quantum link model was investigated in
\cite{Sha04,Ban13,Car17,Hua19,Tsc19,Bro19,Fel20,Car20,Cel20,Luo20,Zac21,Sur20,
Kan20,Hal20,Ban20,Dam21}. Quantum dimer models \cite{Rok88} in condensed matter 
physics have the same Hamiltonian as the $U(1)$ quantum link model on the 
square lattice, but realize the Gauss law in an unconventional manner. They 
also display crystalline confinement and flux fractionalization 
\cite{Ban14a,Ban16}. 

Quantum link models are not limited to these phenomena, but can even be used as 
a regularization of QCD itself \cite{Bro99}. Gluon fields then emerge 
as collective excitations of discrete quantum link variables and quarks arise 
as domain wall fermions. Thanks to the finite-dimensional link Hilbert space of
quantum link models, this alternative formulation of QCD is well-suited for the
implementation in quantum simulation experiments \cite{Wie13,Ban20a}. In 
particular, quarks and gluons can be embodied by ultracold alkaline-earth atoms 
in an optical superlattice \cite{Ban13a}. Quantum simulator constructions for 
$U(1)$ quantum link models with dynamical fermions have used ultracold 
Bose-Fermi mixtures in optical superlattices \cite{Ban12}, while constructions 
without fermions have been based on Rydberg atoms in optical lattices 
\cite{Gla14} or on superconducting quantum circuits \cite{Mar14}. Experimental
digital as well as analog quantum simulations or computations of lattice gauge 
theories including quantum link models have been realized in 
\cite{Mar16,Ber17,Klc18,Lu19,Sch19,Goe19,Mil20,Yan20,Ata21}. The anticipated 
realization of quantum link models in further forthcoming quantum computations 
and quantum simulation experiments motivates the detailed investigation of 
their intricate confinement phases. In this letter, we describe new ``nematic 
confined phases'', in which lattice rotation invariance is spontaneously broken 
while translation invariance remains intact. We also construct a quantum 
circuit that facilitates quantum computations of the corresponding real-time
string dynamics on a chip. 

Let us consider a $U(1)$ quantum link model on a triangular lattice, with a 
2-dimensional link Hilbert space analogous to a quantum spin $\frac12$. The 
two link states carry electric fluxes $\pm \frac12$. The Hamiltonian takes 
the form
\begin{equation}
H\!=\!\sum_\bigtriangleup H_\bigtriangleup\!=\!
- J \sum_\bigtriangleup \left[U_\bigtriangleup + U_\bigtriangleup^\dagger - 
\lambda (U_\bigtriangleup + U_\bigtriangleup^\dagger)^2\right].
\end{equation}
Here $U_\bigtriangleup = U_{xy} U_{yz} U_{zx}$ is an operator associated with the
parallel transport around a triangular plaquette $\bigtriangleup$. It is built
from quantum link operators $U_{xy}$ connecting nearest-neighbor sites $x$ and 
$y$. A $U(1)$ quantum link $U_{xy} = S_{xy}^1 + i S_{xy}^2 = S_{xy}^+$ is a raising
operator of electric flux $E_{xy} = S_{xy}^3$, constructed from a quantum spin 
$\frac12$, $S_{xy}^a$ ($a \in \{1,2,3\})$, associated with the link $xy$. 
The first term in the Hamiltonian inverts a closed loop of electric flux around 
a triangular plaquette. It also annihilates non-flippable plaquette states, 
i.e.\ those that do not contain a closed flux-loop. The Rokhsar-Kivelson term, 
proportional to $\lambda$, counts flippable plaquettes. The Hamiltonian 
commutes with the generators of infinitesimal $U(1)$ gauge transformations, 
which correspond to the lattice divergence of the electric flux operators,
\begin{equation}
G_x = \sum_{i=1,2,3} (E_{x,x+\hat i} - E_{x-\hat i,x}).
\end{equation}
Here $\hat i$ denotes unit-vectors in three lattice directions separated by
120 degree angles. In the absence of external charges, physical states 
$|\Psi\rangle$ obey the Gauss law $G_x |\Psi\rangle = 0$. When static external
charges $Q_x \in \{\pm 1, \pm 2, \pm3\}$ are installed at the lattice sites $x$,
the Gauss law is locally modified to $G_x|\Psi\rangle = Q_x|\Psi\rangle$.
Besides the $U(1)$ gauge symmetry, the model also has several global symmetries,
including lattice translations, rotations, and reflections, as well as charge 
conjugation C, which replaces $U_{xy}$ by $U_{xy}^\dagger$ and $E_{xy}$ by
$- E_{xy}$. We consider a rhombic-shaped lattice volume of side-length $L$ with 
periodic boundary conditions, which is equivalent to a regular hexagon with 
side-length $L/\sqrt{3}$, thus maintaining all lattice symmetries even in a 
finite volume. 
%The torus topology implies an additional global $U(1)^2$ center 
%symmetry associated with ``large'' gauge transformations \cite{tHo79}. The
%corresponding super-selection sectors are characterized by wrapping electric 
%fluxes $F_1 = E_2 - E_3$, $F_2 = E_3 - E_1$, $F_3 = E_1 - E_2$, where
%$E_i = \frac{1}{L} \sum_x E_{x,x+\hat i} \in \Z/2$. The $F_i \in \Z$ commute 
%with the Hamiltonian, but cannot be expressed through ``small'' periodic gauge 
%transformations $G_x$. It should be noted that the three $F_i$ are not
%independent because $F_1 + F_2 + F_3 = 0$.

It is natural to introduce dual degrees of freedom: quantum height variables 
which are associated with the hexagonal lattice that is dual to the original 
triangular lattice. The dual hexagonal lattice is bi-partite and consists of 
two sublattices $A$ and $B$. The height variables on sublattice $A$ are 
associated with the center $\widetilde x$ of an original triangle and take 
values $h_{\widetilde x}^A \in \{0,1\}$, while the height variables on sublattice 
$B$ take the half-integer values $h_{\widetilde x}^B \in \{- \frac12,\frac12\}$. 
A configuration of height variables is associated with a flux configuration
\begin{equation}
\label{electricfield}
E_{x,x+\hat i} = \left(h^A_{\widetilde x} - h^B_{\widetilde x'}\right) \mbox{mod} \, 2
= \pm \frac12.
\end{equation}
Here $\widetilde x = x + \frac{1}{3}(\hat i - \hat j)$ and
$\widetilde x' = x + \frac{1}{3}(\hat i - \hat k)$ where 
$j = (i-1) \mbox{mod} \, 3$ and $k = (i+1) \mbox{mod} \, 3$. It should be noted
that, for a given flux configuration, the height variables are uniquely defined
only up to a global shift $h_{\widetilde x}^X \rightarrow [h_{\widetilde x}^X + 1] 
\mbox{mod} \, 2$ ($X \in \{A,B\}$). The introduction of the dual height 
variables guarantees that the Gauss law of the original flux variables is 
automatically satisfied modulo 2. In order to impose the full Gauss law, the 
height variables are subject to a corresponding constraint. In order to define 
the height variables in the presence of odd charges $Q_x \in \{\pm 1, \pm 3\}$, 
one must connect these charges by Dirac strings running along the links of the 
original triangular lattice. Across a Dirac string, one of the adjacent height 
variables must be shifted by 1 modulo 2.

In order to identify the symmetry breaking patterns in the different phases, we 
introduce two order parameters 
\begin{equation}
M_A = \frac{2}{L^2}
\sum_{\widetilde x \in A} \left(h_{\widetilde x}^A - \frac12\right), \quad
M_B = \frac{2}{L^2} \sum_{\widetilde x \in B} h_{\widetilde x}^B,
\label{orderparameters}
\end{equation}
associated with the two sublattices (each with $L^2$ plaquettes such that
$M_A, M_B \in [-1,1]$). Due to the global shift ambiguity of the height 
variables, $(M_A,M_B)$ and $(- M_A,- M_B)$ are physically equivalent.
It is important to understand the transformation behavior of the order 
parameters under the following symmetries: the charge conjugation C, the 60 
degree rotation O around a point on the triangular lattice, the reflection R 
on a lattice axis, and the reflection R' = R O on an axis orthogonal to a 
lattice axis. The order parameters transform as
\begin{eqnarray}
&&^C M_A = M_A, \quad ^C M_B = - M_B, \nonumber \\
&&^O M_A = M_B, \quad ^O M_B = - M_A, \nonumber \\
&&^R M_A = M_B, \quad ^R M_B = M_A, \nonumber \\
&&^{R'} M_A = M_A, \quad ^{R'} M_B = - M_B. 
\end{eqnarray}

It is straightforward to set up a Euclidean time path integral for the canonical
partition function $Z = \mbox{Tr} [\exp(- \beta H) P]$ (at inverse temperature
$\beta$) using the dual height variable representation. Here the operator $P$, 
which commutes with the Hamiltonian, imposes the Gauss law by projecting onto
the Hilbert space of physical states. We have developed an efficient quantum
Monte Carlo cluster algorithm (cf.\ \cite{Ban18,Ban21}) that operates on the 
height variables, one sublattice at a time. Equal-valued height variables are 
connected to clusters according to rules that guarantee detailed balance. 
Special rules apply in the last time-slice in which the projection operator $P$ 
enforces the Gauss law. The algorithm has been implemented in continuous 
Euclidean time \cite{Bea96}.

\begin{figure}[tbp] 
\vskip-1.2cm
\includegraphics[width=0.235\textwidth]{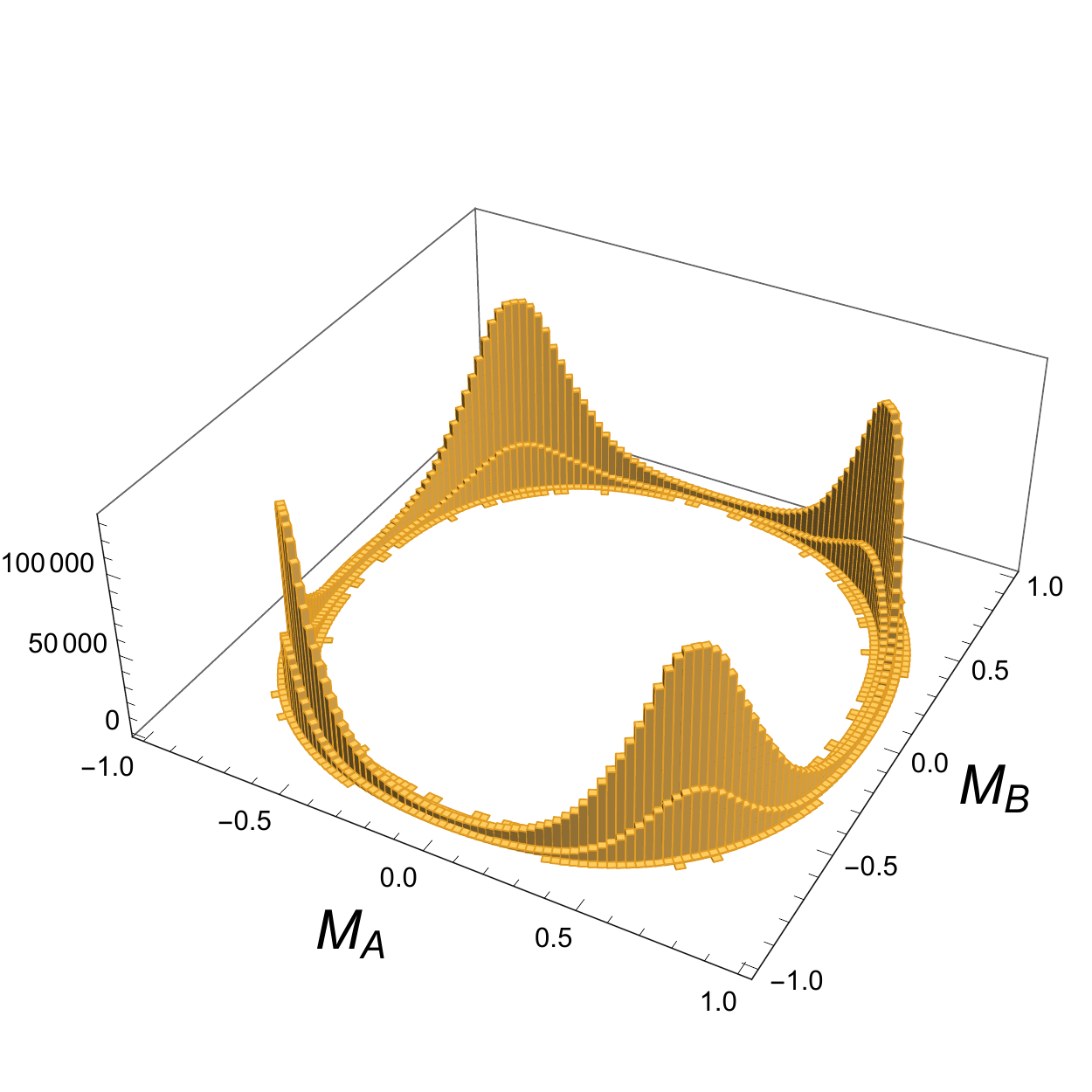} 
\includegraphics[width=0.235\textwidth]{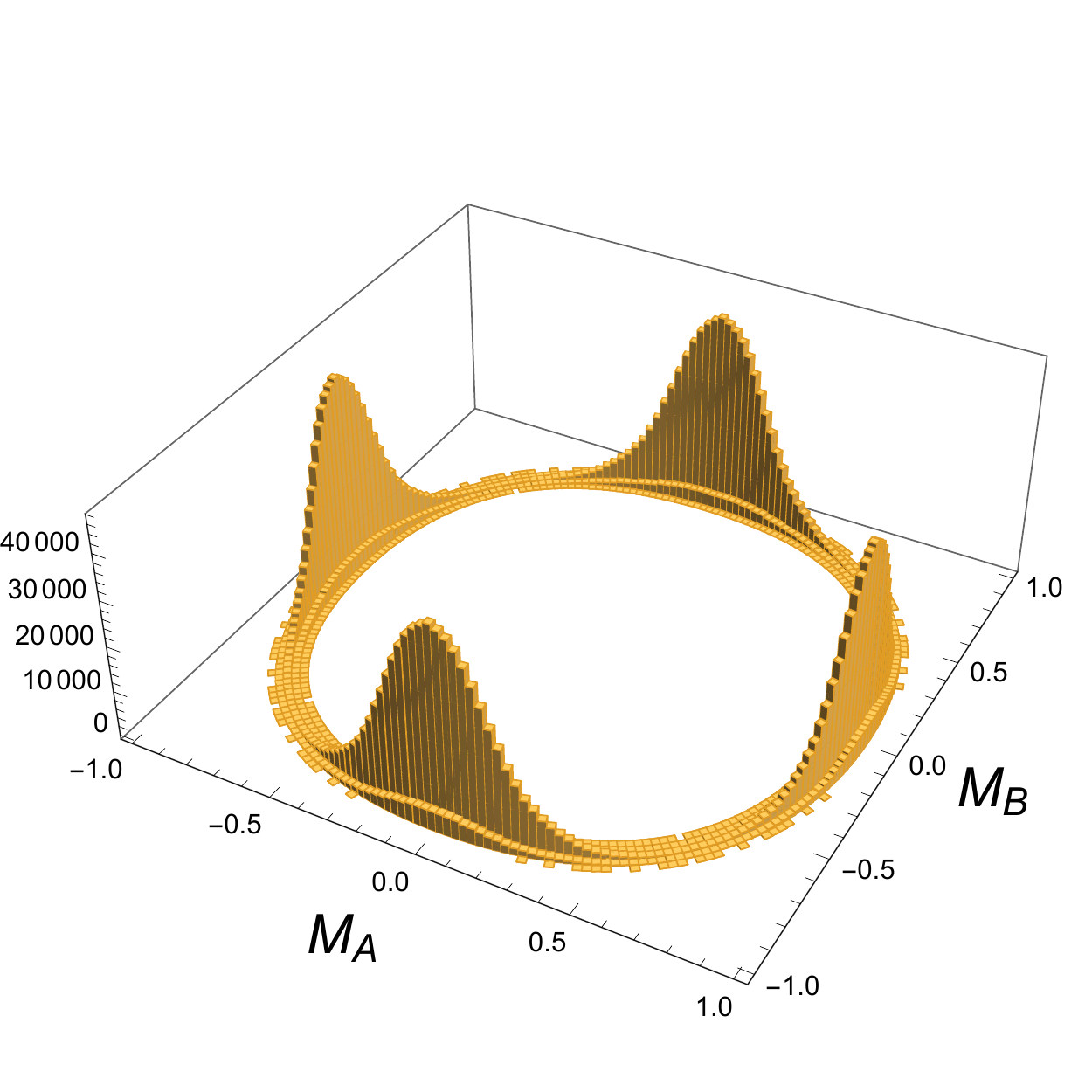} \\ \vskip-1cm
\includegraphics[width=0.235\textwidth]{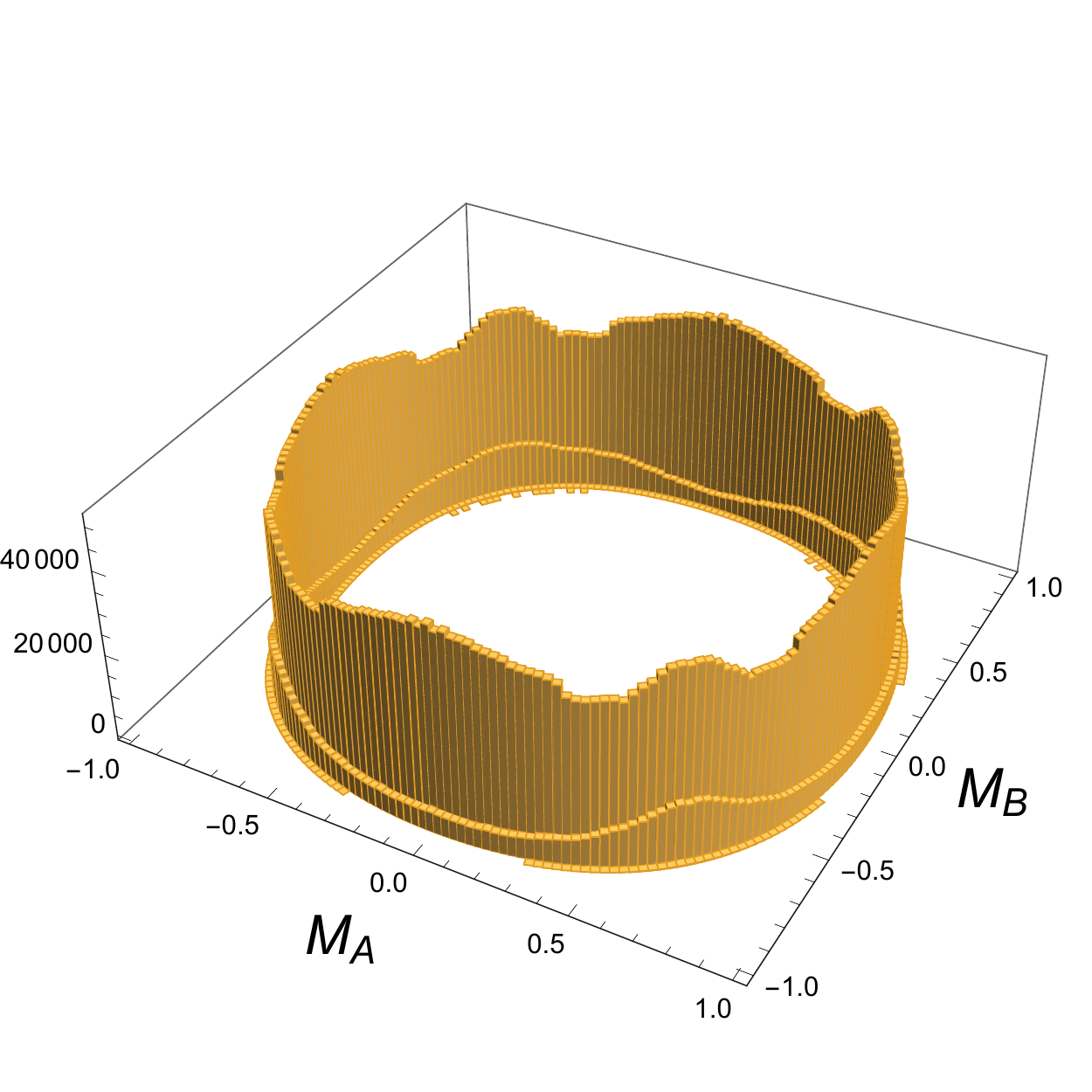}
\includegraphics[width=0.235\textwidth]{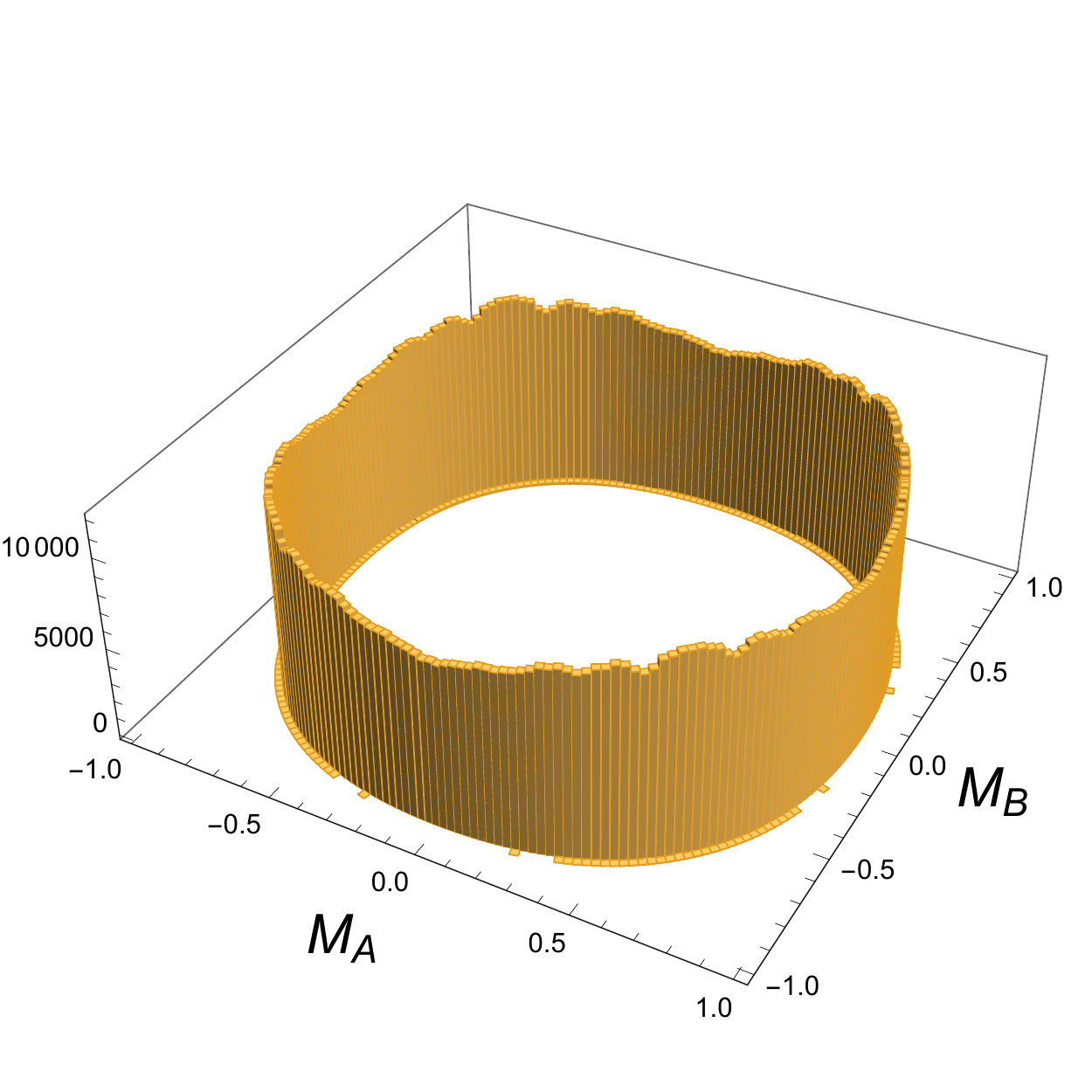}
\caption{[Color online] \textit{Order parameter distributions in the $(M_A,M_B)$
plane for $L = 64$ at $\lambda = -0.2156$ (a), $-0.2146$ (b), $-0.2152$ (c),
and for $L = 48$ at $\lambda = -0.214425$ (d).}}
\label{Fig1}
\end{figure}

In order to explore the phase structure, first in the absence of external 
charges, we have performed Monte Carlo simulations on systems with
$L = 8, 16, 32, 48, 64$ at temperatures corresponding to $\beta J = L$. We 
have explored the region $\lambda \leq 0$ where the cluster algorithm is 
applicable. Fig.\ref{Fig1} shows the probability distribution of the order 
parameters $(M_A,M_B)$ for different values of $\lambda$. For 
$\lambda > \lambda_c = - 0.215(1)$  (Fig.\ref{Fig1}a) the height variables on 
one of the two sublattices order, which implies that the 60 degree rotation 
symmetry $O$ is spontaneously broken. For $\lambda < \lambda_c$ 
(Fig.\ref{Fig1}b), on the other hand, both sublattices order. This implies 
that, in addition to $O$, also the charge conjugation symmetry C is 
spontaneously broken. Since translation invariance remains unbroken in both 
phases, we have encountered two distinct nematic phases.

Remarkably, the phase transition at $\lambda_c$ is associated with an emergent
almost exact spontaneously broken $SO(2)$ symmetry, which manifests itself in 
the ring-shaped order parameter distribution shown in Fig.\ref{Fig1}c. The
corresponding pseudo-Goldstone boson is dual to an almost massless photon. Thus,
the model mimics certain aspects of a deconfined quantum critical point
\cite{Sen04,Sen04a}. However, unlike for deconfined quantum criticality, the 
radius of the ring-shaped order parameter distribution does not shrink to zero 
at the transition. A similar behavior was first observed for the $U(1)$ quantum 
link model on the square lattice \cite{Ban13}, but has also been found in other 
systems \cite{Zha18}. As a result, the transition that separates the two 
distinct nematic phases is an exotic first order phase transition, with an order
parameter that remains large at the transition, while there is still a large 
correlation length due to the almost massless emergent pseudo-Goldstone boson. 
At present, we do not understand the origin of these long-range correlations. 
Weak first order phase transitions have been associated with slowly walking 
couplings near a conformal point \cite{Gor18}. While it would be interesting to 
explore this idea in the context of the $U(1)$ quantum link model, in this 
paper we focus on the corresponding confining string dynamics.

\begin{figure}[tbp]
\vskip-0.5cm
\includegraphics[width=0.5\textwidth]{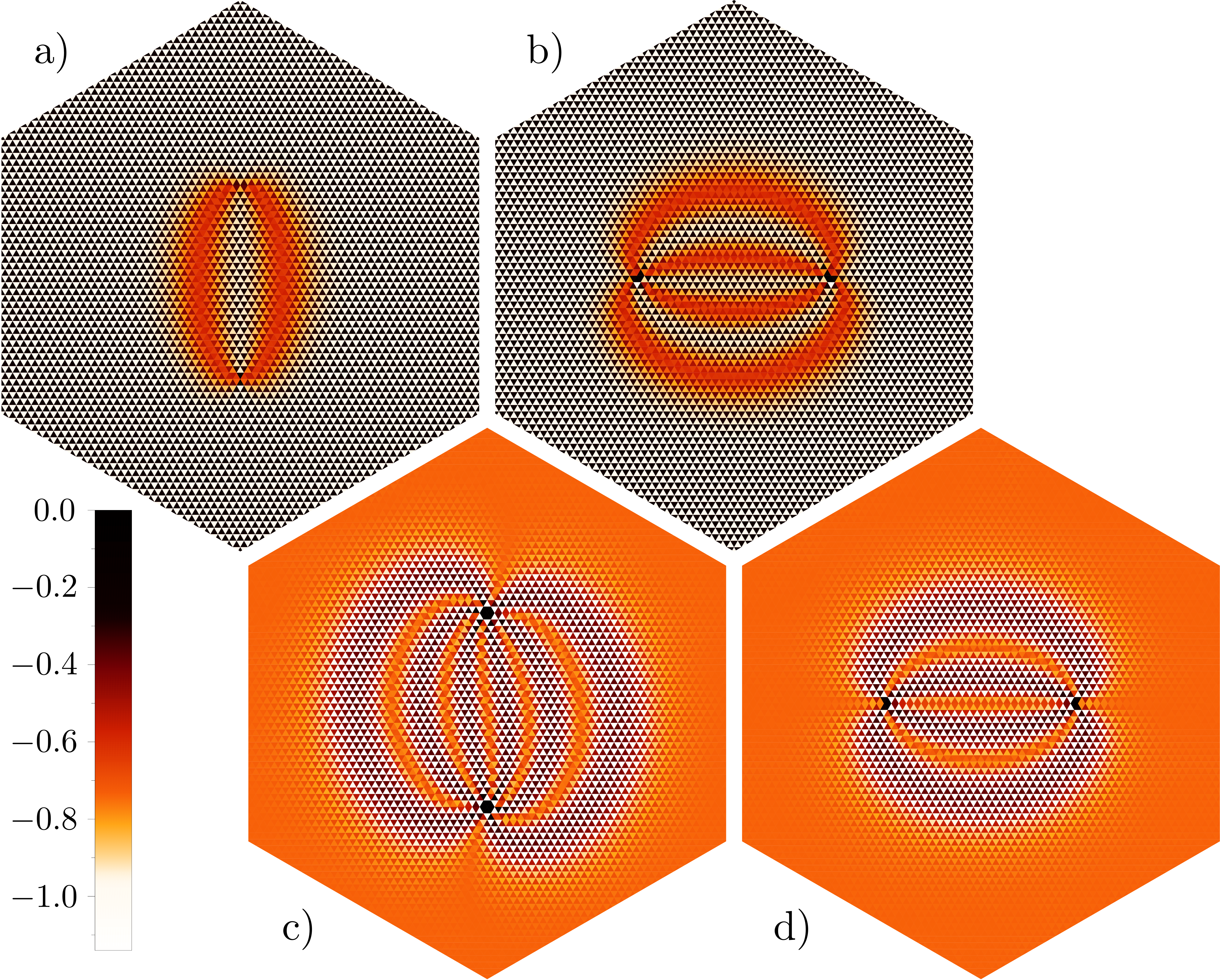} 
\caption{[Color online] \textit{Energy distribution for the strings
connecting two charges $\pm 1$ at distance $r = 15 \sqrt{3}$ (a), and $\pm 2$ 
at $r = 26$ (b), with $\lambda = - 0.1 > \lambda_c$, as well as $\pm 3$ at 
distance $r = 15 \sqrt{3}$ (c), and $\pm 2$ at $r = 26$ (d), with 
$\lambda = -0.3 < \lambda_c$.}}
\label{Fig2}
\end{figure}

We now proceed to the physics in the presence of external charges. We begin 
with the phase at $\lambda > \lambda_c$ in which O but not C is spontaneously 
broken. Fig.\ref{Fig2}a illustrates the energy density of the confining string 
that connects two charges $\pm 1$ separated along a line that is orthogonal to 
a lattice axis. The string separates into two distinct strands, each carrying a 
fractionalized flux $\frac12$. The strands are interfaces that separate the
two degenerate bulk phases. Indeed, in the region between the strands the 
flippable triangular plaquettes are on sublattice $B$, while they are on 
sublattice $A$ in the surrounding bulk. The external charges are responsible for
an explicit breaking of translation invariance, of the charge conjugation C, 
and of the rotation O. There are two types of reflections that are not 
explicitly broken in the presence of the external charges. One is the 
reflection R' on the line connecting the charges. The other is a combination of 
C with the reflection R on the lattice axis that maps the charges onto each 
other. While R' is not spontaneously broken in the bulk, the combination CR is. 
Not unexpectedly, the spontaneous breakdown of CR in the surrounding bulk 
manifests itself in a very slight asymmetry in the strands. Fig.\ref{Fig2}b 
shows the situation with two external charges $\pm 2$ separated along a lattice 
axis. The string then fractionalizes into four strands which separate regions 
of alternating bulk phases. In this case, neither R nor CR' are explicitly 
broken by the external charges, but the spontaneous breakdown of R in the bulk 
is responsible for a visible asymmetry in the strand geometry.

Next, we consider the other nematic phase with $\lambda < \lambda_c$ in which
both O and C are spontaneously broken. Fig.\ref{Fig2}c shows the structure of a 
string connecting two charges $\pm 3$ separated along a line orthogonal to a 
lattice axis. First of all, the string now 
fractionalizes into six strands, which again separate alternating bulk phases. 
Interestingly, the interior of the strands consists of the bulk phase that 
occurs on the other side of the phase transition. As before, for the present 
arrangement of charges the symmetries R' and CR are not explicitly broken. 
However, unlike in the nematic phase with $\lambda > \lambda_c$, for 
$\lambda < \lambda_c$ not only CR but also R' is spontaneously broken in the 
bulk, while the combination CRR' remains unbroken. This explains the symmetry 
of the corresponding strand geometry. Two charges $\pm 2$ separated along a 
lattice axis are shown in Fig.\ref{Fig2}d. As before, in this situation both R 
and CR' are not explicitly broken. However, unlike for $\lambda > \lambda_c$, 
for $\lambda < \lambda_c$ neither R nor CR' is spontaneously broken, which 
explains the reflection symmetries in the corresponding strand geometry.

\begin{figure}[tbp]
\includegraphics[width=0.475\textwidth]{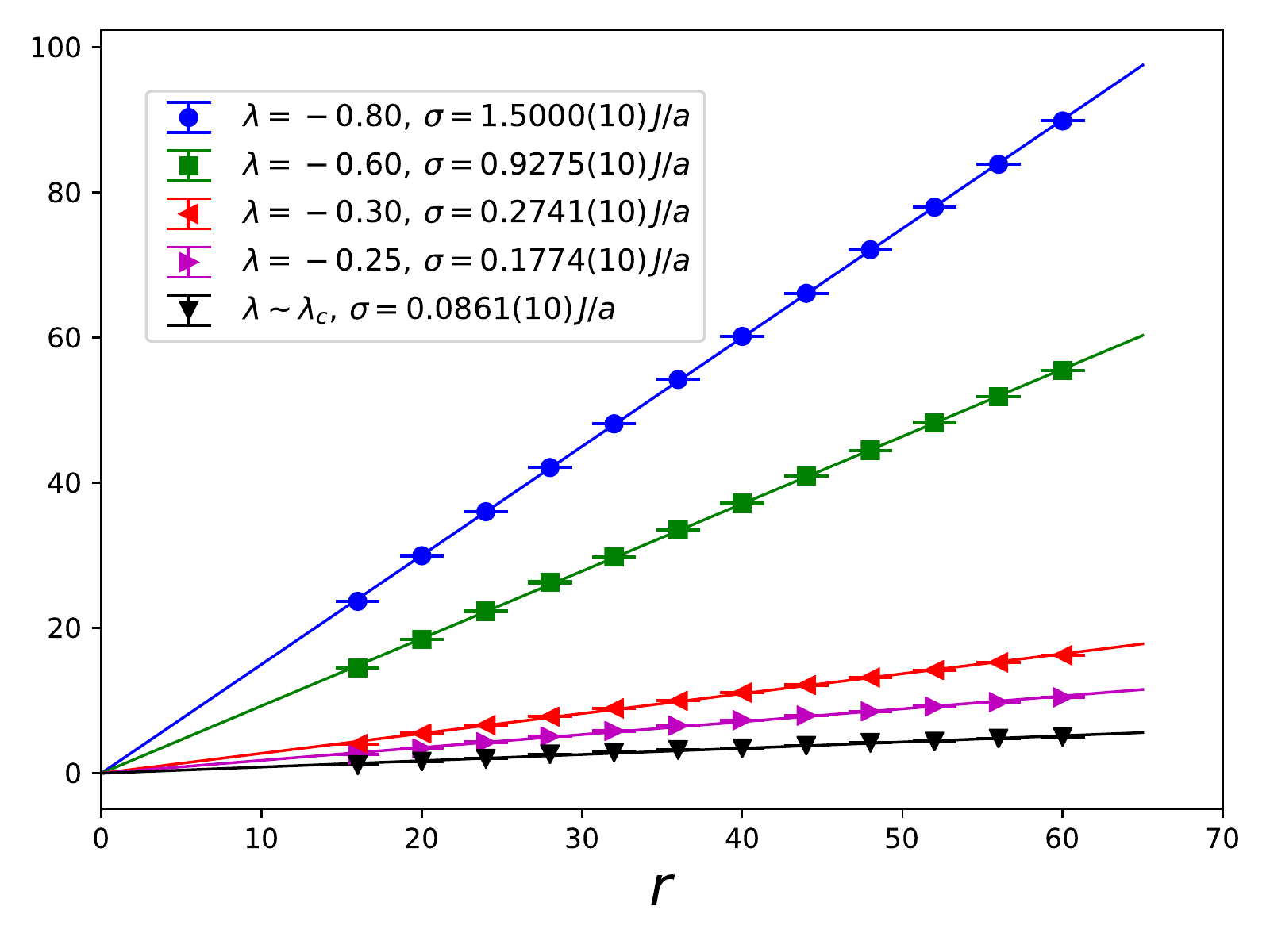}
\caption{[Color online] \textit{Static charge-anti-charge potential $V(r) 
\sim \sigma r$ as a function of the separation $r$ and the corresponding 
string tension $\sigma$ for various values of $\lambda$.}}
\label{Fig3}
\end{figure}

Let us consider the static charge-anti-charge potential $V(r)$ in the
nematic confinement phase with $\lambda < \lambda_c$ for two charges $\pm 2$
separated along a lattice axis, as in Fig.\ref{Fig2}d. As expected,
at large separation $r$ the potential is linearly rising, i.e.\ 
$V(r) \sim \sigma r$ with the string tension $\sigma$ (cf.\ Fig.\ref{Fig3}). As 
one approaches $\lambda_c$, $\sigma$ becomes very small but does not go to 
zero, indicating again that the phase transition is weakly first order. The 
minimal value of $\sigma$ is reached at the transition.

\begin{figure}[tb]
\includegraphics[width=\linewidth]{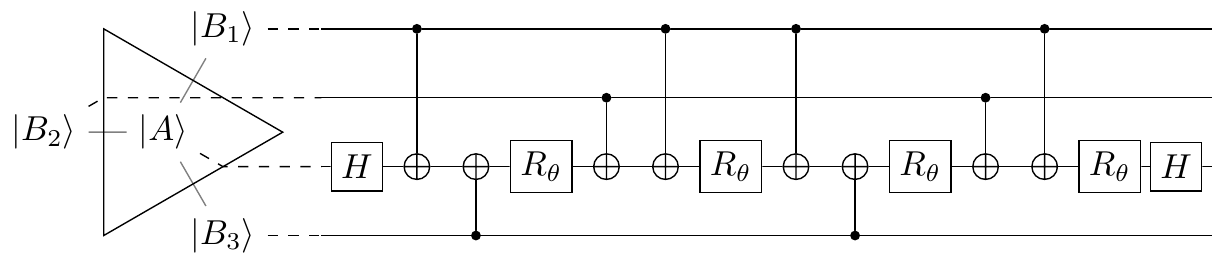}
\caption{\textit{Circuit decomposition of $\exp(- i H_\bigtriangleup t)$ at 
$\lambda = 0$, using two Hadamard gates $H$, four single qubit rotations 
$R_\theta$ with $\theta = - J t/2$ and eight CNOT gates.}}
\label{Fig4}
\end{figure}

Finally, we realize the model as a quantum circuit that embodies the quantum
link model on a chip. The dual formulation can also be used to perform 
real-time quantum simulations. Since the Gauss law constraint is partially 
resolved, this leads to a denser encoding of the physical Hilbert space than 
working with the fluxes $E_{xy} = S_{xy}^3$ themselves \cite{Bro19}. Each dual 
height variable $h^{A,B}$ maps exactly to one qubit 
$|a, b = 0,1 \rangle$ ($a = h^A,\, b = h^B + \frac12$). The electric field of 
eq.(\ref{electricfield}) is diagonal in this basis and is given in terms of the 
adjacent qubit operators as
$E_{x,x+\hat i} = \frac12 Z_{\tilde x}^A Z_{\tilde x'}^B$.
While in the dual formulation there are no single link operators 
$U_{x,x+\hat i}$, operators for closed loops still exist. The plaquette operators 
$U_\bigtriangleup$, e.g., are given by four-qubit operators acting on the 
plaquette qubit $A$ and its three nearest neighbors $B_i$. Exact cancellations 
in the full plaquette Hamiltonian $H_\bigtriangleup$ yield at most three qubit 
interactions
\begin{eqnarray}
H_\bigtriangleup&=&
\frac{J}{4} (1 + Z_{B_1}Z_{B_2} + Z_{B_2}Z_{B_3} + Z_{B_3}Z_{B_1})(\lambda - X_A) 
\nonumber \\
&=&H_0 + H_1 + H_2 + H_3.
\end{eqnarray}
The real-time evolution is using a Trotter scheme which 
alternates between the plaquette terms on the $A$ and $B$ sublattices, while 
fully preserving Gauss's law. Each sublattice Trotter step factorizes into 
mutually commuting single plaquette unitaries $\exp(- i H_\bigtriangleup t) = 
e^{-i H_0 t} e^{-i H_1 t} e^{-i H_2 t} e^{-i H_3 t}$. Each factor translates into a 
very short gate sequence, as illustrated for $\lambda = 0$ in Fig.\ref{Fig4}, 
using eight two-qubit CNOT gates in combination with four single qubit 
rotations $R_\theta = \exp(- i \theta Z/2)$ as well as two Hadamard gates 
$H = (X + Z)/\sqrt{2}$ for each plaquette.

\begin{figure}[tb]
\includegraphics[width=\linewidth]{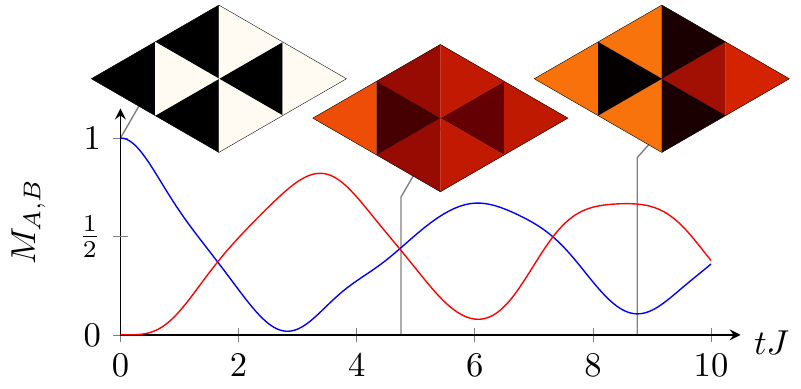}
\caption{[Color online] \textit{Real-time evolution of the order parameters 
$M_{A,B}$ on 8 plaquettes. The energy density is illustrated at three times, 
$tJ = 0, 4.75, 8.75$ in the same way as in Fig.2.}}
\label{Fig5}
\end{figure}

This quantum circuit facilitates quantum computations of the string dynamics on 
a chip. Fig.\ref{Fig5} shows an example of dynamics that can already be studied 
on today's devices. Here the system is a parallelogram of 8 plaquettes with 
fixed boundary conditions $a, b = 0$ on all external plaquettes (not shown in 
Fig.\ref{Fig5}). The system is initialized in the ground state of $H_A$ (the 
sum of all plaquette terms on sublattice $A$ (white triangles)), which happens 
to be a simple product state. The quenched dynamics using the full Hamiltonian 
$H = H_A + H_B$ at $\lambda = 0$ then leads to an oscillation between the order 
parameters $M_{A,B}$ of the two sublattices (cf.\ eq.(\ref{orderparameters})). 
This system also contains a flux string $E = \frac12$ connecting fractional 
charges $\pm \frac12$ at the top and bottom corner. Initially that flux is 
wired along the left boundary, but it starts to oscillate along with the order 
parameters.

Another worthwhile phenomenon to study is the crossing of the quantum phase 
transition in real time. Fig.\ref{Fig2} shows that the interior of the strands 
consists of the bulk phase that is realized on the other side of the 
transition, such that the string interior and the surrounding bulk must 
exchange their roles during this transition.

The rich string dynamics of the simple $U(1)$ quantum link model provides 
additional motivation to push the experimental frontier forward in this 
direction, towards the ultimate goal of quantum simulating QCD itself 
\cite{Wie14}.

\newpage

The research leading to these results has received funding from the 
Schweizerischer Na\-tio\-nal\-fonds and from the European Research Council 
under the European Union's Seventh Framework Programme (FP7/2007-2013)/ ERC 
grant agreement 339220.

\end{document}